\documentclass[twocolumn,english]{aa}
\usepackage[T1]{fontenc}
\usepackage[latin1]{inputenc}
\usepackage{graphicx}
\usepackage{natbib}
\usepackage{babel}
\usepackage{txfonts}

{\rm }

\begin{document}

\title{VLT FORS-1 observations of NGC 6397: Evidence for mass segregation
\thanks{Based on observations collected at the VLT, ESO-Paranal, Chile}}

\author{G. Andreuzzi \inst{1,2}, V. Testa \inst{1}, G. Marconi \inst{1,3},
G. Alcaino \inst{4}, F. Alvarado \inst{4}, R. Buonanno \inst{1,5}}

\offprints{G. Andreuzzi \email{gloria@mporzio.astro.it}}

\institute{INAF-Osservatorio Astronomico di Roma, via Frascati 33, 00040 Monteporzio
Catone, Italy \\
 \and INAF-Centro Galilei Galilei, Santa Cruz de La Palma, Spain\\
 \and ESO-Chile, Alonso de Cordova 4107, Vitacura, Casilla 19001, Santiago, Chile\\
 \and Isaac Newton Institute of Chile, Ministerio de Educacion de
Chile, Casilla 8-9, Correo 9, Santiago, Chile \\
 \and Dipartimento di Fisica, Universit\`a di Roma ``Tor Vergata'', 00100 Roma, Italy }

\date{Received ; accepted }

\authorrunning{G. Andreuzzi et al.} \titlerunning{Mass Segregation in NGC\,6397}

\abstract{We present (V,V-I) VLT-FORS1 observations of the Galactic Globular
Cluster NGC~6397. We derive accurate color--magnitude diagrams and
luminosity functions (LFs) of the cluster Main Sequence (MS) for two
fields extending from a region near the centre of the cluster out
to $\simeq$ 10$\arcmin$. The photometry of these fields produces
a narrow MS extending down to V $\simeq$ 27, much deeper than any
previous ground based study on this system and comparable to previous
HST photometry. The V, V-I CMD also shows a deep white dwarf
cooling sequence locus, contaminated by many field stars and spurious
objects. We concentrate the present work on the analysis of the MSLFs
derived for two annuli at different radial distance from the center
of the cluster. Evidence of a clear-cut correlation between the slope
of the observed LFs before reaching the turn-over, and the radial
position of the observed fields inside the cluster area is found.
We find that the LFs become flatter with decreasing radius (x $\simeq$
0.15 for $1\arcmin<R_{1}<5.5\arcmin$; x$\simeq$ 0.24 for $5.5\arcmin<R_{2}<9.8\arcmin$
; core radius, r$_{c}$= 0.05$\arcmin$), a trend that is consistent
with the interpretation of NGC 6397 as a dynamically relaxed system.
This trend is also evident in the mass function.

}

\maketitle
\section{Introduction}

Globular cluster stars fainter than the Main Sequence Turn-Off (MSTO)
are substantially unevolved and are located along the MS according
to their initial mass. Their luminosity distribution is usually described
by the Initial Mass Function as modified by subsequent dynamical evolution
to become the so-called Present Day Mass Function (PDMF). The precise
knowledge of the Initial Mass Function is a cornestone in a variety
of astrophysical problems, ranging from the physics of stars formation
to the dynamical and chemical modeling of Galactic evolution. In this
respect, GGCs are perhaps the best suited halo structures to investigate
the faint PDMF and Initial Mass Function (IMF), and to verify the
existence and size of a population of very low mass stars, possible
major contributors to the dark halos in galaxies. The mass segregation
then should be seen as a radial dependence of the LFs, but to verify
these theoretical predictions we need to obtain a statistically significative
sample of data starting from the central regions of the cluster. Until
now the only instrument available to obtain a precise luminosity function
of the main sequence below the turn-off at various distances from
the centre of a cluster was the HST. Nowadays, this is also possible
with ground based observations thanks to the large collecting area
and huge resolving power of the VLT. In this work we concentrate
on NGC 6397. There are several reasons for this choice: i) it
is the closest cluster, hence it is possible to reach the faintest
part of the MS; ii) it is a classical, old, metal-poor halo cluster,
and its age constitutes a lower limit to the age of the Galaxy and,
therefore, to the Universe as a whole; iii) an extended study carried
out with HST on a central area over several years
\citep[see][]{cool96,king98} allowed us to separate through proper motions
the cluster from the background/foreground field, and is a valuable
comparison source with our data; iv) a fairly well defined white dwarf
(WD) sequence is visible. \\
 NGC 6397 is located
at $\alpha=17^{h}41^{m}53^{s}$,
$\delta=-53^{\circ}44^{\prime}30^{\prime\prime}$ and has a metallicity
{[}Fe/H{]} = -1.82 \citep{cg97}, slightly higher
than the fiducial most metal-poor cluster M\,92 ({[}Fe/H{]} = -2.15
\citep{grat97}. High-quality populated color--magnitude diagrams
(CMD) have been produced in the past either from the ground
\citep[see, e.g.,][]{alca87,alca97} and from the space \citep{cool96,dema95a}.
The distance has been recently re-determined by \cite{reid98}
with a set of extreme subdwarfs from the catalogue
of HIPPARCOS, obtaining $(m-M)_{0}=12.13\pm0.15$. The reddening adopted
by the authors, which is a weighted mean of the available estimates
is $E(B-V)=0.18$. We will adopt these values in the following. The
field of view covered by our data ranges over a 8.8$^{\prime}$ radius, starting
at about 1$^{\prime}$ from the cluster center. This displacement
allowed us to study the presence of mass segregation in the luminosity
function on NGC \,6397. The structure of the paper is as follows:
Sects. \ref{sect:obs} and \ref{sect:analysis} present the data,
the reduction procedures and the V, V-I CMD. The resulting LF and MF
are presented and discussed in Sects. \ref{sect:funzioni} and
\ref{sect:mf}. Sect. \ref{sect:discussion} summarizes the results.

\section{Observations and data reduction}

\label{sect:obs}

The observations consist of two fields (f1 and f2 in the text), including
several 600 s exposures in each of the two V and I Bessel filters,
taken with the standard resolution of the VLT + FORS1 (0.2 $\arcsec$/pixel)
corresponding to a field of view of 6.8$\arcmin$ $\times$ 6.8 $\arcmin$
(see Table \ref{tab:tab}). The data taken in service mode in 1999
(f2) and in 2000 (f1) were retrieved electronically from the ESO-STECF
archive (proposals: 63.H-0721(A) and 65.H-0531(B)).

\begin{table}

\caption{Journal of observations for the fields f1 and f2. N$_{I}$ and N$_{V}$
are the number of exposures respectively for the filter I and V; t
is the exposure time in seconds.}

\label{tab:tab}

\[
\begin{array}{c|cc|cc}
\mathrm{field} & \alpha & \delta & \mathrm{N_{I}\times\mathrm{t}} & \mathrm{N_{V}} \times t\\
\hline f1 & 17^{h} 41^{m} 14.3^{s} & -53^{0} 44' 27.6'' & 59\times600 & 42\times600\\
f2 & 17^{h} 41^{m} 03.5^{s} & -53^{0} 45' 36'' & 11\times600 & 13\times600\end{array}\]

\end{table}

\begin{figure}

\caption{The location of the HST-fields \citep{cool96} inside
the ground-fields. The position of the center of the
cluster is (6.71 $\arcmin$, 8.5 $\arcmin$). Dashed circles represent
the annuli in which we divided our sample to perform data analysis (see
text for further details)}
\resizebox{\hsize}{!}{\includegraphics{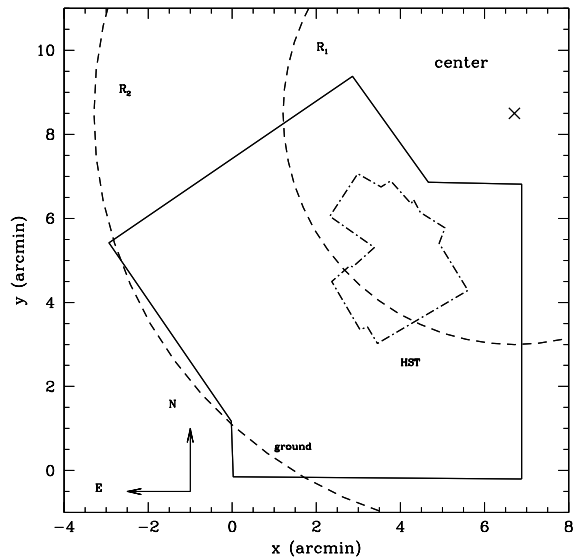}}

\label{fig:mappa}
\end{figure}

The two partially overlapped fields are located at $\simeq$ 5$\arcmin$
SE from the cluster center and cover a region of the
cluster already observed with the HST-WFPC2 by \cite{cool96}
(see Fig. \ref{fig:mappa}). The presence of a region in common between
the two ground fields and the HST fields allowed us to obtain a
homogeneous calibration for the whole ground-sample.

Corrections to the raw data for bias, dark and flat-fielding were
performed using the standard IRAF routines following the same recipe
employed in the ESO-VLT pipeline. Subsequent data reduction and analysis
was done by following the same procedure for both the data-sets, and
using the DAOPHOT-ALLFRAME package \citep[Version3]{stet87} with
a quadratically varying point spread function (PSF).

Since our goal was to reliably detect the faintest objects in both
fields, first we created two median images using all available
frames in each filter of a given field, then we added these median
images together to obtain the final images on which to search for objects
by running the standard DAOPHOT routines. The
candidates were then measured on each of the individual V and I frames
and an average instrumental magnitude was derived for those recovered
in at least three frames.

As our fields were affected by a large number of saturated stars
and, as the photometric procedures
can include those peaks in the final photometric catalogue, we decided
to eliminate from the catalogue the spurious identifications before
performing the data analysis. In this way, a total of 15081 and 11161
objects were detected, respectively, in the fields f1 and f2.

We used the 7059 objects in common between these two fields to optimize
our photometry by computing a weighted average between the values
of the magnitudes of a given star in each catalogue and to transform
the frame coordinates to a common coordinate system referred to the
field f2. At the end of this procedure an homogeneous set of instrumental
magnitudes, color and position were obtained for a total of 19228
objects.

Conversion from instrumental magnitudes to the HST VEGAMAG standard
system, which is similar to the Johnson system \citep{holtz95}
was performed by using the 3559 objects in common between the two
ground data sets and the catalogue from HST observations \citep{cool96}.

\section{The Color Magnitude Diagram and the Luminosity Function}

\label{sect:analysis}

\begin{figure}

\caption{V, V-I color-magnitude diagram for 16105 objects selected from the
original catalogue ($\sigma_{I}$, $\sigma_{V}\leq0.15$). The mean
errors per interval of magnitude V are also plotted.}
\includegraphics[width=10cm]{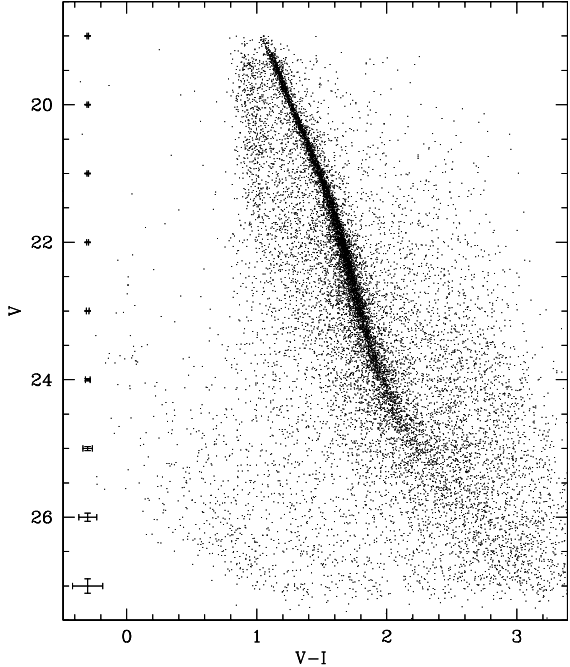}

\label{fig:cmdtot}
\end{figure}

Figure \ref{fig:cmdtot} shows the V, V-I CMD for 16105 objects of the
original catalogue for which photometric errors are: $\sigma_{I},\sigma_{V}\leq0.15$.

This figure shows the high quality of our photometric
data. A very narrow MS (0.028 $\leq\sigma(V-I)\leq$
0.15 for 19 $\leq$ V $\leq$ 25), extending down to V $\sim$ 27 is visible,
much deeper than any previous ground based study \citep{alca87,alca97}
and comparable only with the most recent HST photometry
\citep{cool96}. From the figure a deep
white dwarfs cooling sequence locus is also clear, although contaminated at the
faintest magnitudes by many field stars and possibly spurious objects.

\subsection{The white dwarf cooling sequence}

\begin{figure}

\caption{Comparison of NGC\,6397 cooling sequence with hydrogen-rich WD
cooling sequences (solid lines) of 0.5, 0.7 and 0.9 M$_{\odot}$,
obtained by using models from Hansen (1998, 1999).}
\resizebox{\hsize}{!}{\includegraphics{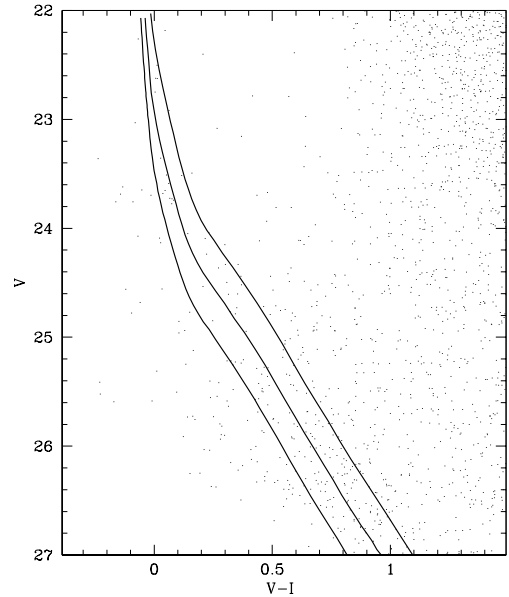}}

\label{fig:cooling}
\end{figure}

In Fig. \ref{fig:cooling} we overplot  theoretical cooling sequences
for 0.5, 0.7 and 0.9 M$_{\odot}$ hydrogen-rich WDs
on the CMD. Theoretical cooling sequences have been obtained using the WD cooling
models of Hansen et al. (1998, 1999) as described in Richer et al. (2000) and
have been trasformed to apparent magnitudes and colors
appropriate for NGC\,6397 adopting  $\mathrm{(m-M)_{0}}=12.13$ and $\mathrm{E(B-V)}=0.18$.

Since the population of WDs accumulates at lower luminosities as the age of the
cluster increases, we could fix a limit to this age by looking
at the location of the faintest WDs in the cooling sequence.
However the high contamination by field stars in this region of the CMD does not allows 
us to extract precise information.
By looking at the faint end of the cooling sequence (V $\simeq$ 27), 
the age of the oldest WDs in the cluster ranges from $\simeq$ 2 Gyr to $\simeq$ 3.8 Gyr,
for WDs with mass ranging from 0.5  M$_{\odot}$ to 0.9  M$_{\odot}$ respectively.

\subsection{The Luminosity Function}

Before using the CMD to build the MSLF, we carefully selected the
sample in order to keep only the \emph{bona fide} real objects.
In particular we found that they appear in at least 80 $\%$ of the
original frames. We adopted the following selection procedure:
for each field, we first built two maps (the 'weight-maps') with
the same dimension of the median images I and V used for candidate searching
(see sect. \ref{sect:obs} for details). These maps are such that
the value of each pixel is the number of times the pixel appears on
the stack, i.e. how many frames are stacked onto each other on that pixel,
and ranges from {[}1 - N$_{max}${]}, where N$_{max}$ is the total number of frames
overlapped in the original median image.
With this information we found, for each object, the number
of frames in which it has been recovered with respect to the number of
frames in which it is expected to be detected.

\begin{figure*}

\caption{V, V-I color-magnitudes diagram obtained selecting the original catalogue
for different values of the ratio $r = r_{I} = r_{V}$.} 
\centering
\includegraphics[width=13cm]{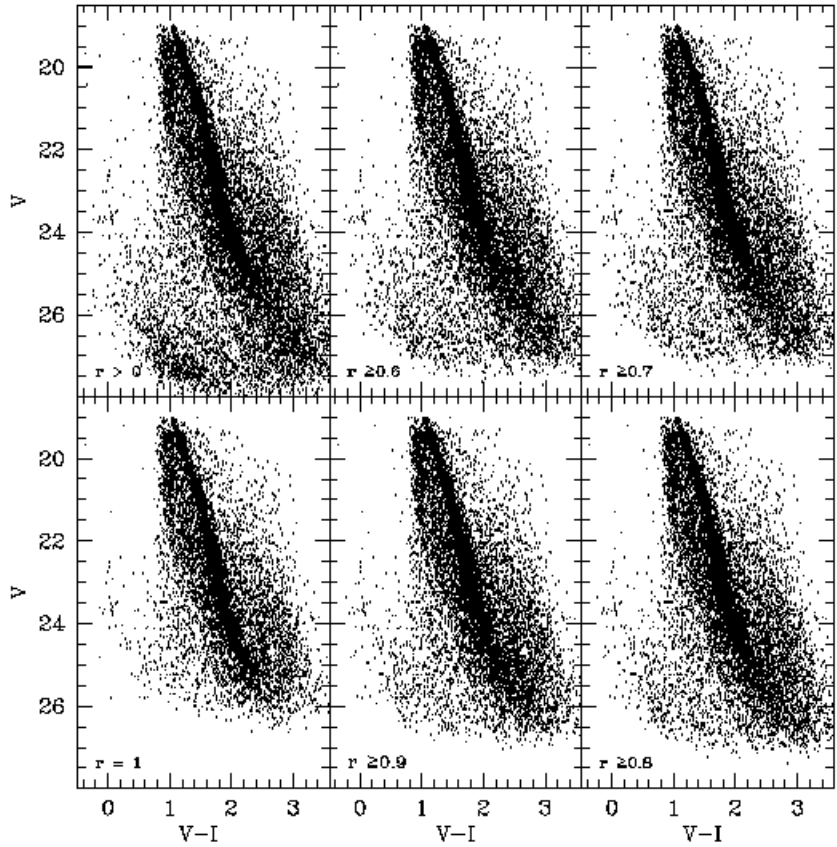}

\label{fig:cmd}
\end{figure*}

\begin{figure}

\caption{LFs obtained selecting the original photometric catalogue for different
values of the ratio $r = r_{I} = r_{V}$.}
\resizebox{\hsize}{!}{\includegraphics{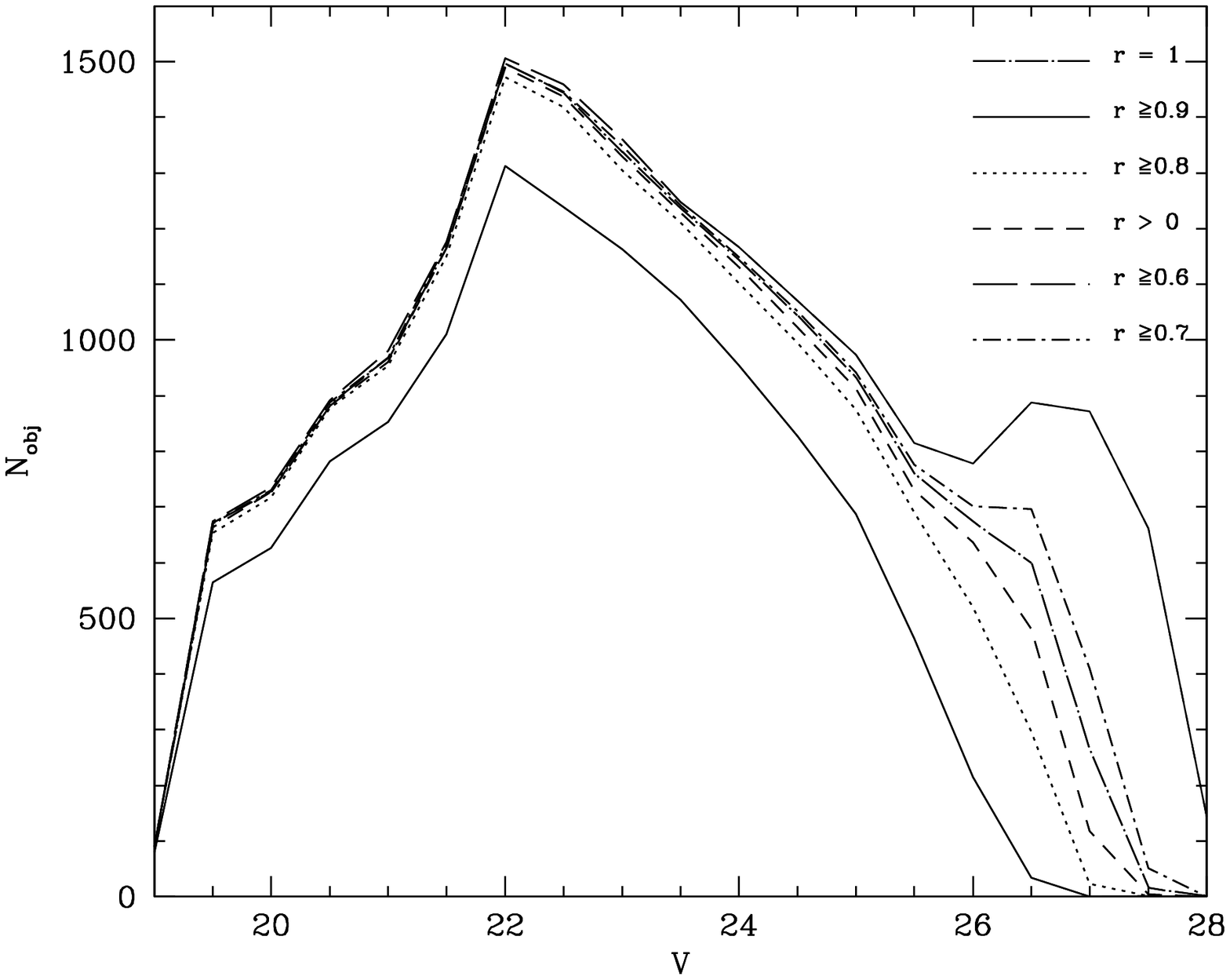}}

\label{fig:fl}
\end{figure}

Figure \ref{fig:cmd} and \ref{fig:fl} show the variation of the CMD,
and the correspondent LF,
with the parameters $r_{I}$ and $r_{V}$ defined, respectively, as the ratio
between the number of frames in which a star has been recovered ($n_{I,V}(rec)$)
and the number of the original frames available for that pixel ($n_{I,V}(frame)$):

\begin{equation}
r_{I,V}=\frac{{\displaystyle n_{I,V}(rec)}}{{\displaystyle n_{I,V}(frame)}}\end{equation}

In particular, for the two cases: $r_{I,V}>$ 0, and $r_{I,V}$ = 1,
we are looking at: i) all the objects of the original catalogue (without
any selection); ii) only the objects recovered simultaneously in all
the original frames. 


\begin{figure}

\caption{Comparison between the LFs obtained from a catalogue in which the objects
are selected by following two different criteria: photometric errors
$\leq$ 0.15 (open triangles); number of frames in which a star has
been recovered (larger than 80 $\%$) (dashed line).} 
\resizebox{\hsize}{!}{\includegraphics{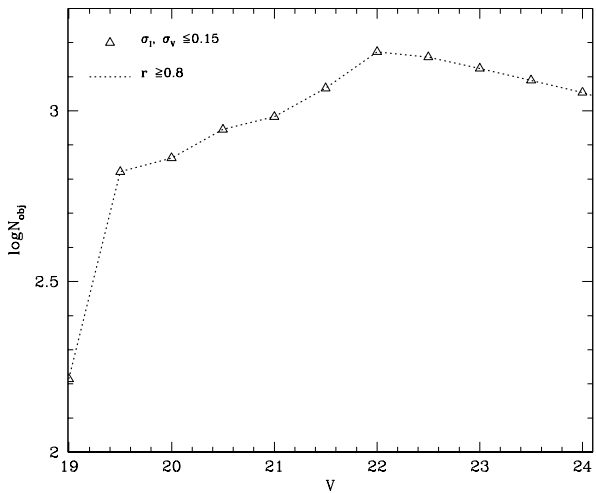}}

\label{fig:er}
\end{figure}

An analysis of these figure shows that, by increasing the value of the
parameter r, and thus decreasing the number of the selected objects
in the catalogue: i) the distribution of the objects located in the
region of the CMD we are interested, i.e. the MS stars with a magnitude
brighter than V = 22, (the peak of the LF), does not change; ii) we
mainly lose information about the faint region of the CMD and in particular
for the objects located at the faint end of the cooling sequence (see
the case r $>$ 0), where the photometric errors are very large ($\sigma_{I},\sigma_{V}\geq0.3$)
; iii) for a choice of the parameter r $\geq$ 0.8 (objects of the
original catalogue recovered in at least 80 $\%$ of the original
frames), the CMD is very similar to that shown in Fig. \ref{fig:cmd}.
This last suggestion may be confirmed by the comparison between the
LFs associated with the CMDs reported in  Fig. \ref{fig:er}, 
where open triangles represent the LF obtained from
a catalogue selected for $\sigma_{I},\sigma_{V}\leq$ 0.15 and dashed
line is the LF from a catalogue selected for $r\geq0.8$. 

\section{The Main Sequence Luminosity Function}

\label{sect:funzioni}

\subsection{Sample selection and completeness}

\begin{figure}

\caption{V, V-I color-magnitude diagram for our data-sample. Solid line on
the MS represents the ridge line; solid lines outside the MS delimit
the region of the CMD that we will use to build the MSLF.}
\resizebox{\hsize}{!}{\includegraphics{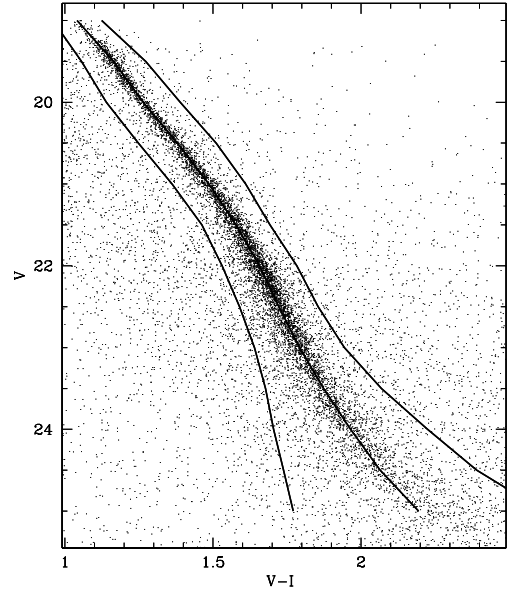}}

\label{fig:ridge}
\end{figure}

Since our goal in this work is to study the MSLF, in the following
we will concentrate our analysis only on the region of the CMD including
the MS stars. In order to select the objects located on the MS locus,
we built a fiducial line assuming a Gaussian color distribution,
then we determined the mean color and dispersion $\sigma$, in each
magnitude interval. We accepted as MS stars only the objects for which
the distance from the ridge line is $\leq$ 3 $\sigma$. Figure \ref{fig:ridge}
shows the MS locus selected with this procedure. The procedure is very robust where the number of MS stars 
is large, while it is more uncertain below V$\sim$24 where the MS is less populated and comparable to the 
field contribution. However, our analysis stops at brighter magnitudes (see below) and hence the uncertainty
at the faintest magnitude bins has not been taken into account. The 3-$\sigma$ selection around the MS locus 
ensures that almost all the MS contribution is considered while minimizing the contamination from the field.
This would not affect significantly the slope of the LF but would provoke an increase in the errors. Several tests 
have been done with different values of sigma clipping around the MS locus, and the final 3-$\sigma$ is the optimal choice.

Observed MSLFs have been created by dividing the MS in bins 0.5
mag--wide, and counting the stars in each bin, correcting for  
incompleteness. 

The incompleteness depends on the level of crowding in the observed
fields and, therefore, on their location with respect to the cluster
centre. In particular, an insufficient or inappropriate correction
for crowding would result in the distortion of the stellar LF for the
loss or overcorrection of faint star counts. 
In our case, crowding is not the only source
of incompleteness: the distribution in luminosity of the stars is
also modified by the large number of hot pixels and bad columns affecting
the original images. 

The fraction of objects lost in each magnitude bin as a function
of the distance from the center of the cluster has been quantitatively
estimated by using the widely tested artifical star method. It
consists of adding a number of artificial stars to the original frames 
and reducing once again the `artificial frames' by adopting the same 
technique already adopted for the original ones.

As we were interested in verifying the photometric completeness of MS
stars, we simulated artificial objects with magnitude and color appropriate
for the MS. In doing this, the CMD has been divided in
0.5 V-mag bins, and for each bin the mean V and
I magnitudes (derived from the MS ridge line) have been computed.
Then a set of artificial stars have been simulated having the selected
magnitude values.

Artificial stars were first added randomly to the reference I frame, 
paying attention to add only a few percent ($\leq$ 10 $\%$) of the 
total number of stars actually present in the frames in the magnitude
bin, in order to avoid a significant enhancement of the image
crowding. The same stars have then been added to the other I and V
frames in the same position with the respect to the reference frame.

All pairs of V and I `artificial' frames ($\simeq$ 4000 images for
$\simeq$ 428543 artificial stars) were then analyzed in the same way
as the original ones and a final catalogue of artificial
stars has been created for each bin and compared with
the catalogue of the input artificial stars.

An artificial star was considered recovered if the output magnitudes fell in the original
bin and 
$\delta{\textrm{y}}$ $\leq$ 1.5 pixel and $\delta{\textrm{mag}}$
$\leq$ 0.25 mag. If $N_{\textrm{rec}}$ is the number of the recovered
stars and $N_{\textrm{sim}}$ is the number of simulated stars, the
ratio $N_{\textrm{rec}}/N_{\textrm{sim}}=\Phi$, gives the completeness in that bin 
for the location considered. 
With this approach we have built a map showing how photometric
completeness varies with position in the frames. 

In addition to correcting for photometric incompleteness, a reliable
determination of the cluster LF requires a correction for the
contamination caused by field stars. In our case, we are interested
only in the change of the slope of the cluster LF with the distance
from the center of the cluster. Since the field
star contribution is not dependent on this distance, and appears
quite homogeneous over the CMD locus considered \citep[see][]{cool96},
we decided not to correct for this effect, that, however, would
increase the global error on the LF.

\subsection{The Luminosity Function}

\begin{figure}

\caption{Left: V, V-I color-magnitude diagram for stars in the annular rings
S1 (left panel) and S2 (right panel).}
\resizebox{\hsize}{!}{\includegraphics{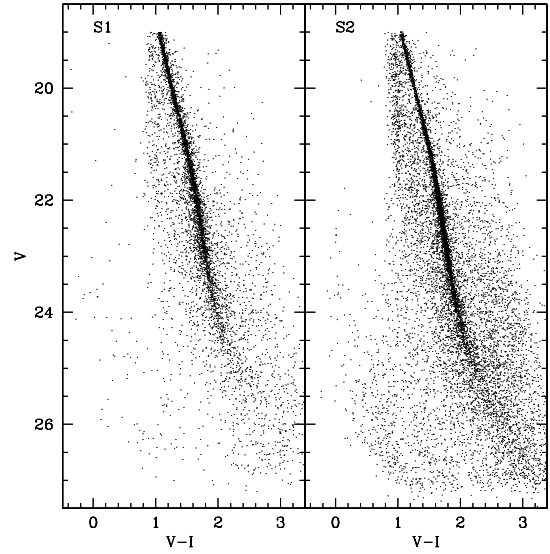}}

\label{fig:cmd12}
\end{figure}

Figure \ref{fig:cmd12} shows the V, V-I CMDs obtained for stars belonging
to the two annular rings S1 and S2 located respectively at 1 $\arcmin\leq R_{1}\leq5.5\arcmin$
and 5.5$\arcmin\leq R_{2}\leq9.8\arcmin$ from the center of the cluster
(see figure \ref{fig:mappa} for details). 
These two regions have been selected in order to divide the sample roughly in half and to keep most of the
HST sample in the inner region. Only a small fraction of the area covered by \cite{cool96} has been left in the
outer annulus for calibration purposes. The S1/S2 radius is also roughly equivalent to $2\times r_h$. In this way
we are quite confident that the population in S1 is representative of the half-mass radius population.

\begin{figure}

\caption{Observed luminosity functions for the two fields S1 (bottom panel)
and S2 (top panel) in bins of 0.5 mag before (dashed line) and after
(solid line) incompleteness corrections; error bars include both Poisson
error and the uncertainty due to the correction for the incompleteness.
The limit of the completeness at the magnitude I = 22.5 is also indicated
for both the fields.}  
\includegraphics[width=10cm]{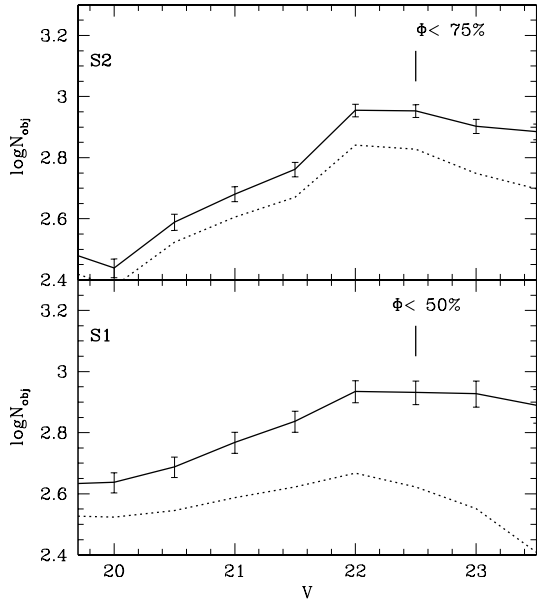}

\label{fig:fl12}
\end{figure}

Figure \ref{fig:fl12} shows the corresponding LFs before (dashed
line) and after (solid line) corrections for incompleteness have been
applied. In the figure error bars reflect the total error associated
with each bin and include both the error on the counts and the uncertainty
due to the correction for incompleteness, which is the main
contributor to the global error figure:

\begin{equation}
\sigma^{2}\approx\frac{{\displaystyle N}}{{\displaystyle \Phi^{2}}}+\frac{{\displaystyle (1-\Phi)N^{2}}}{{\displaystyle N_{sim}\Phi^{3}}}\end{equation}

where N is the number of observed stars in each bin; $\Phi$ is the
incompleteness factor and N$_{\textrm{sim}}$ is the number of simulated
stars in the bin.

This uncertainty has been estimated by
assuming that the number counts follow a Poisson
distribution, and the uncertainty in determining $\Phi$ is
derived from a binomial distribution as shown in Bolte (1989):

\begin{equation}
\sigma_{N}^{2}=N\end{equation}

\begin{equation}
\sigma_{\Phi}^{2}\approx\frac{{\displaystyle \Phi(1-\Phi)}}{{\displaystyle N_{sim}}}\end{equation}

The actual counts, the completeness factor and the counts corrected
for this factor for both the fields S1 and S2 are listed in Table
\ref{tab:compl}. 

\begin{table}

\caption{\label{tab:compl} \baselineskip 0.4cm Luminosity functions for
the fields S1 and S2 in the V band in 0.5 mag bins. N is the actual
number of stars; $\Phi$ is the completeness factor.} \[
\protect\begin{array}{c|rr|rr}
\multicolumn{1}{c|}{} & \multicolumn{2}{|c|}{\textrm{S1}} & \multicolumn{2}{|c}{\textrm{S2}}\protect\\
\cline{1-5}{\textrm{V}} & {\textrm{N}} & {\textrm{$\Phi$}} & {\textrm{N}} & {\textrm{$\Phi$}}\protect\\
\cline{1-5}19 & 129 & 0.81 & 90 & 0.87\protect\\
19.5 & 338 & 0.79 & 279 & 0.87\protect\\
20 & 334 & 0.77 & 239 & 0.87\protect\\
20.5 & 351 & 0.72 & 334 & 0.86\protect\\
21 & 387 & 0.66 & 403 & 0.84\protect\\
21.5 & 419 & 0.61 & 468 & 0.81\protect\\
22 & 465 & 0.54 & 694 & 0.77\protect\\
22.5 & 419 & 0.49 & 673 & 0.75\protect\\
23 & 356 & 0.42 & 560 & 0.70\protect\\
23.5 & 256 & 0.33 & 499 & 0.65\protect\\
24 & 208 & 0.22 & 486 & 0.59\protect\\
24.5 & 153 & 0 & 466 & 0.51\protect\\
25 & 39 & 0 & 179 & 0.43\protect\\
\end{array} \]

\end{table}

By looking at the table we can see that: a) the completeness factor
depends on the distance from the center of the cluster as we expect
when the crowding contribution in a field is important; b) the completeness
degree never exceeds 87$\%$, also at the brightest magnitudes of
the more external field, S2. This last result is a consequence of
the area lost by saturated pixels.
This hypothesis is confirmed from the values found
for area lost in the regions S1 (saturated
area $\sim$ 22 $\%$) and S2 (saturated area $\sim$
9$\%$). 

The analysis of the LFs shows that they present the same characteristics
already found in other galactic globular clusters with a maximum at
V $\simeq$ 22 and a slow decrease to fainter luminosities \citep[ and
  refs. therein]{dema95a,dema95b,piotto97,marcon98,andr00}.
It is, therefore, natural to investigate
whether the effect of a radial dependence of the LF, already observed
in several globular clusters \citep[see, for instance][]{andr00}, 
is found in the present set of data, that covers a significant
interval of radial distances. 

\begin{figure}

\caption{Observed luminosity functions for the two fields S1 (solid line)
and S2 (dashed line) in bins of 0.5 mag, normalized to the peak (V
= 22). Error bars include both Poisson error and the uncertainty due
to the correction for the incompleteness. } \centering
\includegraphics[width=10cm]{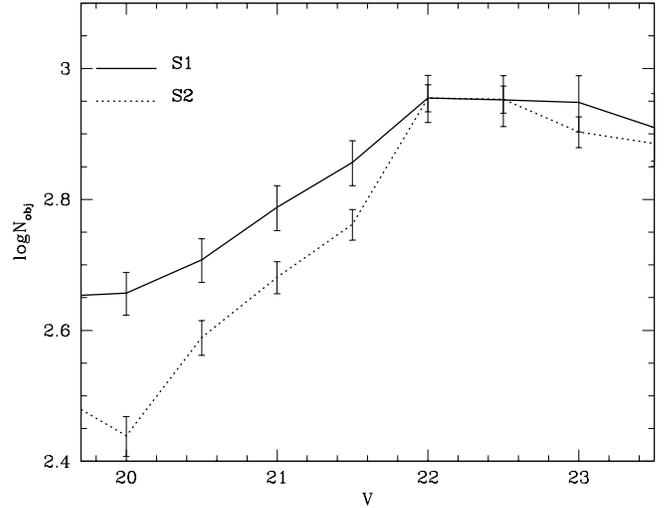}

\label{fig:pic}
\end{figure}

The LFs for S1 and S2, after applying corrections for incompleteness
and normalized to the number of stars in the peak (V = 22), are shown
in Fig. \ref{fig:pic}. 

Inspection of this figure reveals that the two LFs have similar shapes,
but different slopes, in the sense that the ratio between
the bright and the faint stars is larger in S1 than in S2, in qualitative
agreement with the dynamical modifications that stars in globular
clusters experience, as a consequence of the energy equipartition
due to two-body relaxation \citep{spitzer87}. 

\begin{figure}

\caption{Comparison between the LF of S1 (continuous line), corrected for incompleteness,
and the LF of HST (dotted line). The two LFs are normalized to their respective areas. }
 
\resizebox{\hsize}{!}{\includegraphics{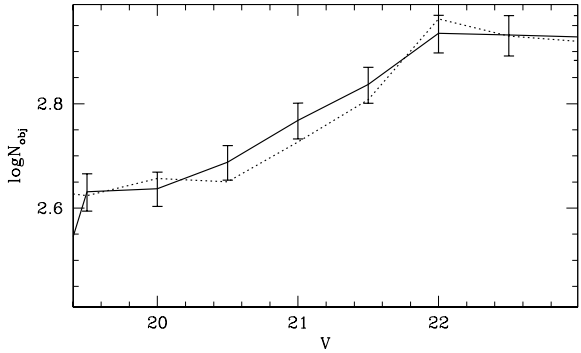}}

\label{fig:comphst}
\end{figure}

To confirm this suggestion we applied a linear fit to the data and
found that in the range in magnitude 20 $\leq$ V $\leq$ 22, the
LF for stars in S2 is steeper (x = 0.24 $\pm$ 0.02) than that for
stars in S1 (x = 0.15 $\pm$ 0.07). 
By taking the slopes with their formal error, this difference 
is marginally significant. A K-S test made on the two distributions
gives a probability of the two distributions to be different of
95\%, i.e. a two-sigma level. However, we compared our S1 LF after
correction for incompleteness with LF obtained with HST, that is much
less affected by incompleteness, and found a good agreement in
the slope. The LF of S1 and HST are reported in Fig. \ref{fig:comphst} 
where the LF of S1 is shown with a continuous line and HST with a dotted line. 
As will be shown below, the MF slopes derived with HST
data and with S1 LF, by using the same set of models, also are in excellent
agreement. These two considerations lead us to conclude that the
difference in the slopes of the LF of S1 and S2 are actually
significant beyond the real meaning of the formal errors. 

\section{The Mass Function}

\label{sect:mf}

\begin{figure}

\caption{The two theoretical relations used to convert luminosities into masses. Short dashed line 
is for Montalban et al. (2000), while long dashed line is for Baraffe et al. (1997).}
\resizebox{\hsize}{!}{\includegraphics[width=8cm]{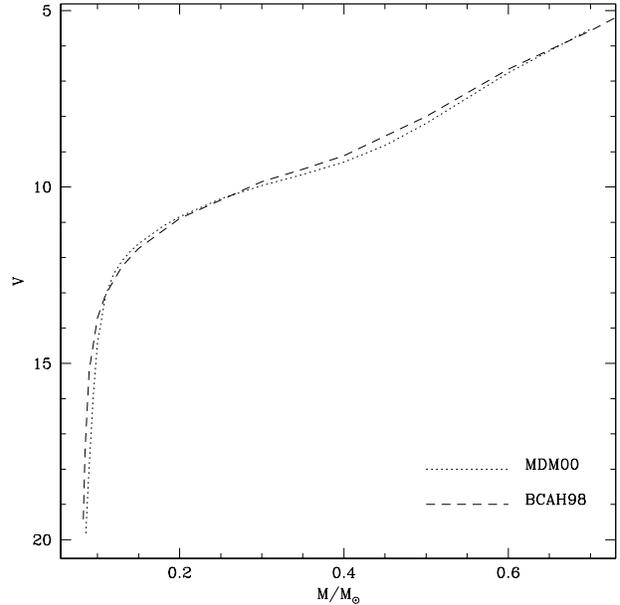}}

\label{fig:theo}
\end{figure}

In order to translate the luminosities into masses and estimate the
PDMF in the two annuli, we used the M/L
relations of \cite{mdm} and \cite{bar97}. The two relations are known
to be similar for our mass range ($M<0.6M_{\odot}$), as pointed out
by \cite{mdm} in their work (see their Fig. 20). 
Small differences are found in the range corresponding to the magnitude
range in which our LF is drawn ($19<V<23$). Figure \ref{fig:theo} shows the two
theoretical relations in the plane (Mass-V), where V is the absolute V magnitude.

We transformed our luminosities into masses for the two annuli, both
models and two metallicites: ${\textrm{[Fe/H]}} = -1.5$ and
${\textrm{[Fe/H]}} = -2.0$,
since the metallicity of NGC 6397 is estimated between these two
values.
Then, we fit a mass function of the form $dN/dM\propto M^{-\alpha}$.
A recent discussion on the form
of the mass function can be found in \cite{dema00}.

\begin{figure*}
\centering
\caption{The MF of S1 and S2 for [Fe/H] = -1.5. The dotted line with
  open squares refers to \cite{mdm}, the dashed with block squares to
  \cite{bar97}. The thicker continuous lines are the fitted relations.}
\includegraphics[width=12cm]{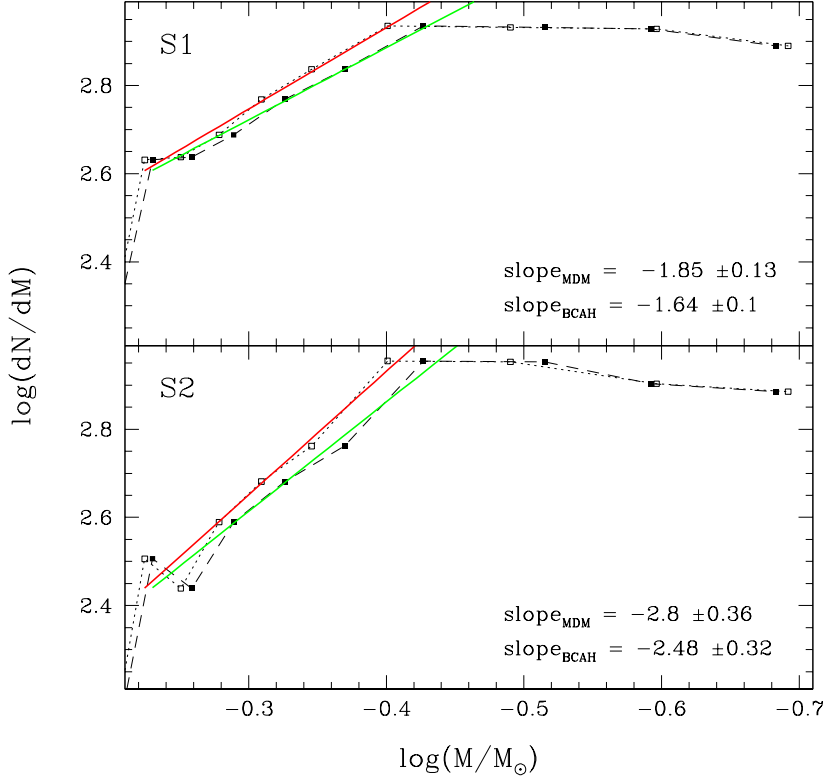}
\label{fig:mf15}

\end{figure*}

\begin{figure*}
\centering
\caption{The MF of S1 and S2 for [Fe/H] = -2.0. Symbols are as in
  figure \ref{fig:mf15}.}
\includegraphics[width=12cm]{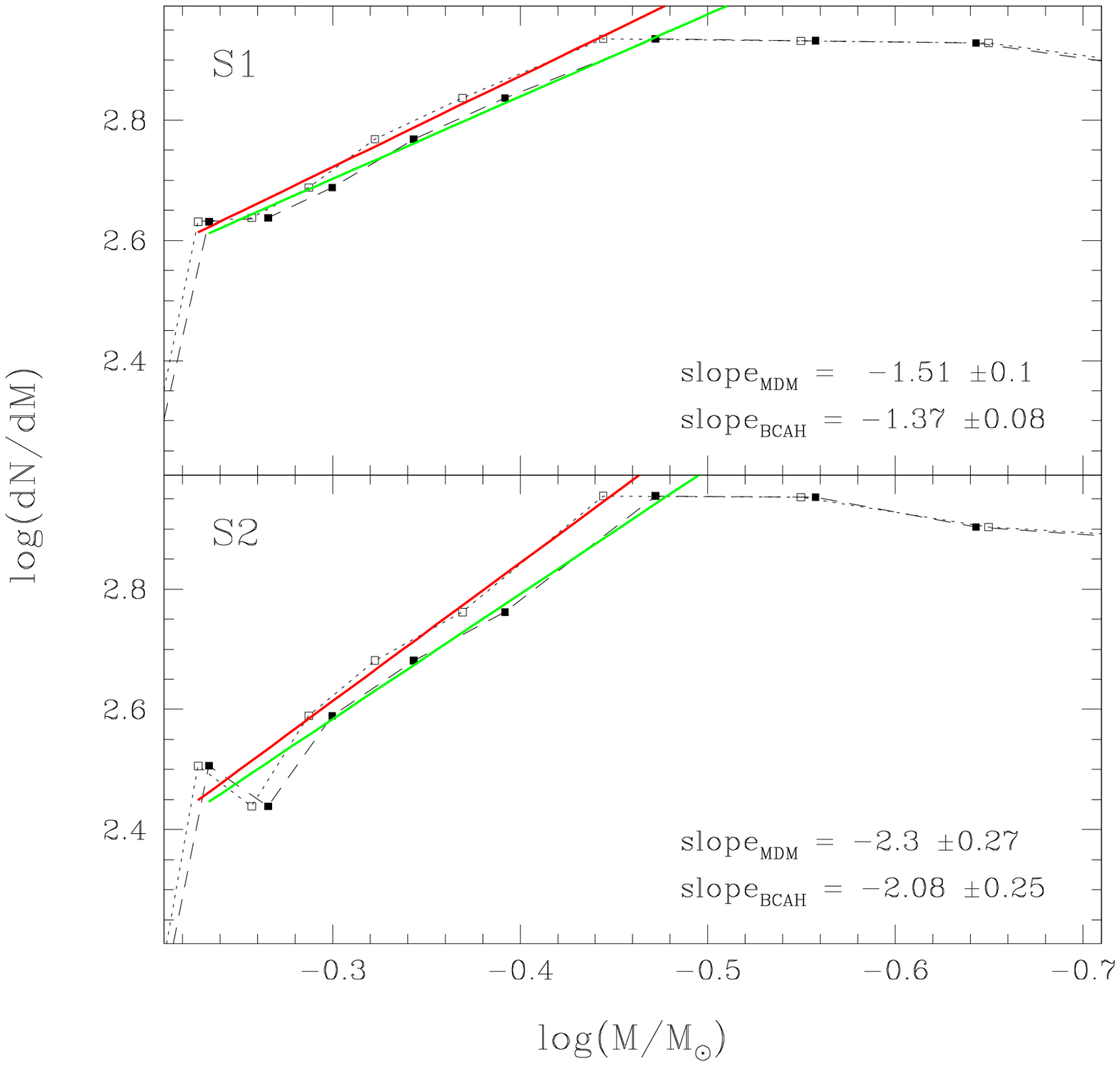}
\label{fig:mf20}
\end{figure*}

Figures \ref{fig:mf15} and \ref{fig:mf20} show the PDMF for the two annuli and the fitting relation,
for the two cited metallicities. The main result is that, also for
the masses, the slopes differ in the two annuli S1 and
S2, for both model sets. The formal uncertainties seems to indicate
that the significance of this difference is high.
A difference in the slope between \cite{mdm}
and \cite{bar97} is also found for both metallicites, \cite{bar97}
being flatter. However, this difference, given the uncertainties on
the fitted slopes, is only marginally significant.

We compared our slope with the one obtained by \cite{mdm} in their
work using the LF from \cite{king98} for the case ${\textrm{[Fe/H]}}=-1.5$.
Our derived slope for the annulus S1, that overlaps the area covered
by \cite{king98} (see Sect. \ref{sect:obs}), is in very good agreement
with the estimate of \cite{mdm} ($\alpha=x+1=-1.5$
versus $\alpha=-1.51$). This is a further confirmation that the completeness
correction to the LF of annulus S1, that is more uncertain, has
been correctly estimated.

\cite{dema00} studied the properties of the PDMF of NGC 6397
by using a data-set obtained with HST-NICMOS and archival
existing WFPC2 (V, I) data and analyzed the radial behavior
of the mass function in the near-IR at two different radii, namely 3.2$^{\prime}$ and
4.5$^{\prime}$ from the cluster center, and in the optical bands out
to 10$^{\prime}$. HST data go considerably deeper in
magnitude while keeping a high sample  completeness. 
\cite{dema00} are able to determine the MF slope down to mass values
of 0.09 $\mathrm{M}_{\odot}$, finding a change in the slope at 0.3
$\mathrm{M}_{\odot}$, that becomes 0.2 $\pm$ 0.1. 
The VLT data-set, especially in the annulus S1,
can only describe the MF down to $V = 22.5$ corresponding to $M \sim
0.32 M_{\odot}$. However, The MF of annulus S2 has a high degree of
completeness down to $V \sim 23.5$ which corresponds to $M \sim 0.2
M_{\odot}$. For this subsample, it is possible to detect the change in
slope found by \cite{dema00}. The slope after the change is $\alpha \sim 0.27
$, with the models of \cite{mdm} and $\alpha \sim 0.30$ with
the models of \cite{bar97}.
Since our completeness correction is
larger than the HST sample, and thus the slope more uncertain, we can
regard the agreement as good.

A further comparison can be made with M\,92 \citep{andr00}, for which
a similar analysis has been performed. In units of $r_{\textrm{h}}$,
the authors cover a similar interval of radial distance and found a
steepening of the MF before the peak, with values comparable to the
ones found for NGC\,6397, if homogeneous models \citep{bar97} are used.
Perhaps a trend in the sense that M\,92 has a shallower overall LF can
be seen and attributed to the difference in metallicity, but recent
results of \cite{pulo03} discuss the possibility that the number of
primordial binaries could play a r\^ole in shaping the MF, thus making
the picture more complicated.

%
%


\subsection{The impact of binaries on the LF of NGC\,6397}

\begin{figure*}
\centering
\caption{LFs of S1 (long dashed line) and S2 (solid line) compared with
         the LF obtained by adding to the MFS2 different percentages
         of binary stars (see text for details). }
\includegraphics[width=12cm]{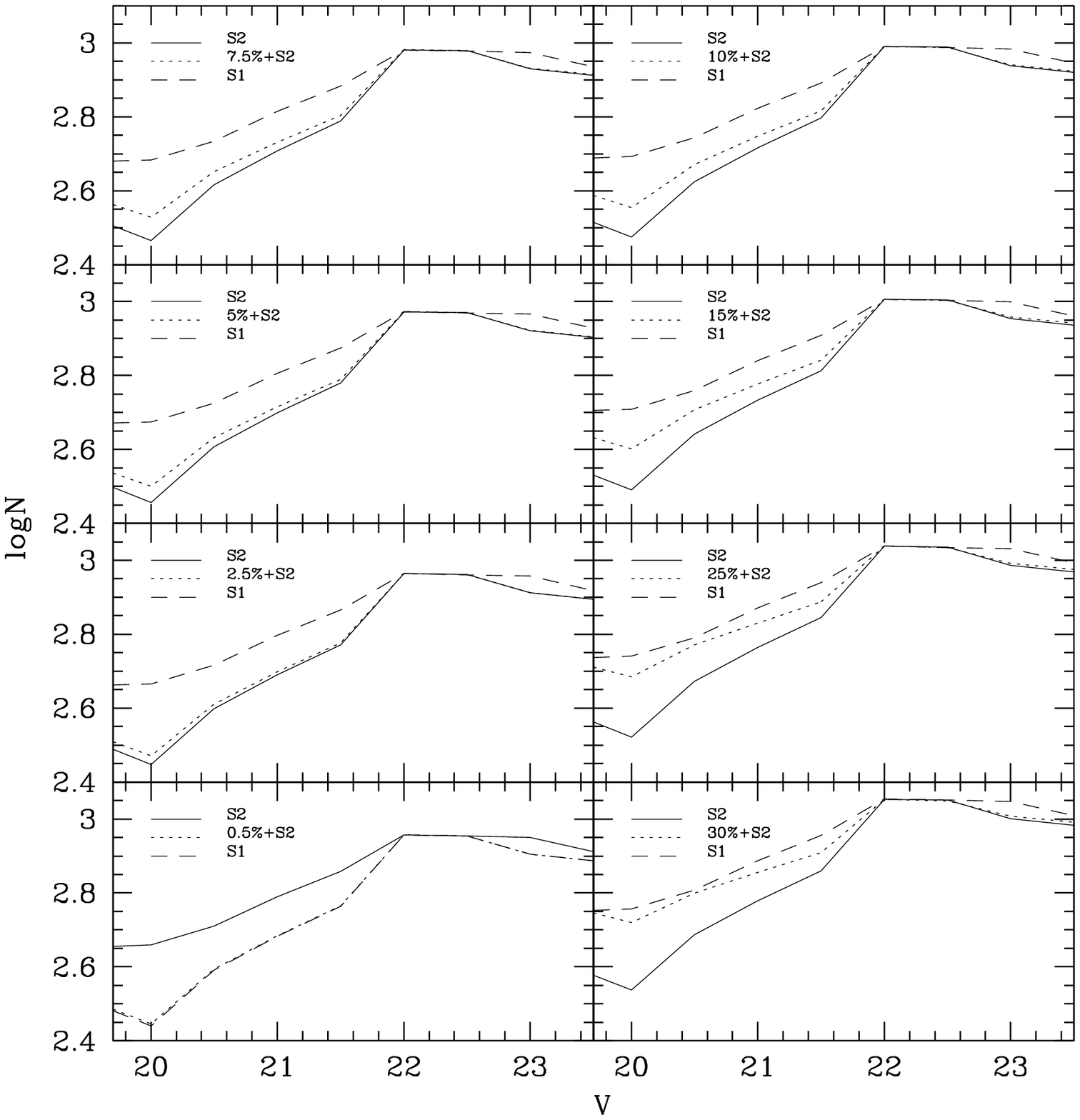}
\label{fig:binarie}

\end{figure*}

In order to check the possible impact of binaries on the LF
of NGC\,6397, we performed the following test: first, we simulated a percentage of
binary stars in each bin in mass of the MF of the
annulus S2, under the simplified assumptions that the annulus S2 is made of single stars and
that binaries are made of two equal mass stars. This is likely an overestimate of the effect, 
but the hypothesis we want to test is that, even with the maximum possible effect, binaries 
alone cannot explain
the difference in the slopes. Then, by using backward the mass-luminosity relation of
Montalban et al. (2000), we calculated the effect of the binaries on the V-magnitude LF of S2.
The corrected LF has been compared with the original LFs of the annuli S1 and S2 as
shown in Fig. \ref{fig:binarie}.
The eight panels of the figure show a comparison between the
LF S1 (long dashed line), LF S2 (solid line) and the LF obtained
by adding to the objects in the LF S2 different fractions of binaries (short dashed lines). 
The LF S1 and LF S2 have been normalized to the peak (V = 22) of the corrected S2 LF.

As already noted by Holtzman et al. (1998), binaries
have a strong influence on the luminosity function of faint stars so
that models with no binaries overestimate the number of faint
stars with respect to models with binaries. 

An analysis of Fig. \ref{fig:binarie} shows that by increasing the percentage of 
binaries the slope of the LF becomes flatter and tends to reach the same 
slope as LF S1. However this happens for a fraction of binaries 
larger than the values predicted by the theory, $\sim$ 25 $\%$. For this value the slope
of the contaminated LF is still steeper ($\sim$ 0.17) than the LF S1 ($\sim$ 0.15). 
If we also consider that the effects of mass segregation inside the cluster
make binaries sink into the central region, 
we are confident that: 1) binaries alone cannot explain the difference in slope
of the two LFs; 2) the sinking of binaries into the cluster central regions has to be regarded as 
an effect of mass segregation too.

\section{Summary}

\label{sect:discussion}

We analyzed the cluster NGC 6397 with data obtained with the VLT and covering
a relatively wide area spanning a radial distance from the cluster
center $1^{\prime} < r < 9.8^{\prime}$. The CMD has been constructed 
and the MSLF obtained after correcting for the incompleteness due to
the crowding. The radial properties of the LF have been studied as
well as the MF after applying mass-luminosity relations from recently
published models. The main results can be summarized as follows:
\begin{itemize}
\item{} The CMD extends down to $V \sim 26.5$, reaching magnitudes
  comparable with previous HST studies. 
\item{} The two annuli in which the sample has been divided show
  different degrees of completeness with respect to the LF. This comes
  from the fact the internal annulus (S1) is close to the cluster
  center and hence its crowding conditions are more critical. This
  prevents us from building the LF fainter than $V = 22.5$. The outer
  annulus (S2) is, instead, complete down to $V \sim 24$. However,
  comparison with the HST sample of \cite{cool96}, that mostly
  overlaps S1, shows that our completeness is consistent. 
\item{} The two annular LFs have been fitted with an exponential law
  obtaining significantly different slopes, down to the last reliable
  magnitude bin, i.e. where completeness drops below 50\% (see
  previouis item). The slope of S1 is flatter than S2, indicating the
  presence of different mass distributions at the two different radial
  distances. This is a clear mark of mass segregation, as expected in
  dynamically relaxed systems such as GCs. 
  The possible effect of binaries has been discussed, concluding
  that binaries alone are unlikely to explain the slope difference 
  between S1 and S2 LFs. 
\item{} To further clarify the situation, we applied two
  different mass-luminosity relations from the models of \cite{mdm}
  and \cite{bar97}, for two different metallicities, ${\textrm{[Fe/H]}} = -1.5$
  and ${\textrm{[Fe/H]}} = -2.0$. The two models, although very similar, give
  slightly different slopes. Both, however, confirm the difference in
  slopes between S1 and S2 also in the MF. A change in slope at $M
  \sim 0.3M_{\odot}$, described in \cite{dema00}, could only be
  studied in the more complete annulus S2, where it has been found
  consistent with the values given by \cite{dema00}. 
\item{} The main source of uncertainty in deriving the LF is given by the sample
  incompleteness that, for ground based studies, is often a strong
  effect also when using an instrument like VLT in good seeing
  conditions. However, part of this effect has been balanced by the
  much higher statistics provided by the large field of view of FORS1,
  9 times greater than that of previous HST studies.
\end{itemize}

\begin{acknowledgements}
We warmly thank Giampaolo Piotto for kindly providing us HST-data for comparison
and data calibrations and Francesca D'Antona for helpful discussions
on the mass-luminosity relations. This work has been supported by the MURST/Cofin2000
under the project: Stellar observables of cosmological relevance. 
\end{acknowledgements}


\begin{thebibliography}{}
\bibitem[\protect\astroncite{Alcaino et al.}{1987}]{alca87}Alcaino, G., Buonanno, R., Caloi, V., et al. 1987, AJ 94, 917 
\bibitem[\protect\astroncite{Alcaino et al.}{1997}]{alca97}Alcaino,
  G., Liller, W., Alvarado, F., Kravtsov, V., Ipatov, A., Samus, N.,
  Smirnov, O., 1997, AJ, 114 1067 
\bibitem[\protect\astroncite{Andreuzzi et
    al.}{2000}]{andr00}Andreuzzi, G., Buonanno, R., Fusi Pecci, F.,
    Iannicola, G., Marconi, G., 2000, A\&A, 353, 944
\bibitem[\protect\astroncite{Baraffe et al.}{1997}]{bar97} Baraffe, I., Chabrier, G., Allard, F., Hauschildt, P., 1997, A\&A, 327, 1054
\bibitem[\protect\astroncite{Bolte}{1989}]{bolte89}Bolte, M. 1989, ApJ 341, 168 
\bibitem[\protect\astroncite{Carretta \& Gratton}{1997}]{cg97}Carretta, E., Gratton, R., G., 1997 A$\&$AS, 121, 95 
\bibitem[\protect\astroncite{Cool et al.}{1996}]{cool96}Cool, A., M., Piotto, G., King, I., R., 1996 ApJ, 468, 655 
\bibitem[\protect\astroncite{De Marchi et al.}{1995a}]{dema95a}De Marchi, G., Paresce, F., 1995 a, A$\&$A 304, 202 
\bibitem[\protect\astroncite{De Marchi et al.}{1995b}]{dema95b}De Marchi, G., Paresce, F., 1995 b, A$\&$A 304, 212 
\bibitem[\protect\astroncite{De Marchi et al.}{2000}]{dema00}De Marchi, G., Paresce, F., Pulone, L., 2000, ApJ, 530, 342
\bibitem[\protect\astroncite{Gratton et al.}{1997}]{grat97} Gratton,
  R. G., Fusi Pecci F., Carretta E. et al. 1997, AJ 491, 749 
\bibitem[\protect\astroncite{Hansen et al.}{1998}]{hans98} Hansen, B., M., S. $\&$ Phinney, E., S. 1998, MNRAS,
  294, 557
\bibitem[\protect\astroncite{Hansen et al.}{1999}]{hans99} Hansen, B., M., S. 1999, ApJ,
  520, 680
\bibitem[\protect\astroncite{Holtzman et al.}{1995}]{holtz95}Holtzman, J., A., Hester, J. J.,  Casertano,
 S., et al., 1995, PASP, 107, 156
\bibitem[\protect\astroncite{Holtzman et al.}{1998}]{holtz98} Holtzman, J., A.,
Watson, A., M., Baum, W., A., et al., 1998, AJ 115, 1946
\bibitem[\protect\astroncite{King et al.}{1998}]{king98}King, I., R. Anderson, J., Cool, A., M., Piotto, G. 1998 ApJ 492L,
37 
\bibitem[\protect\astroncite{Marconi et al.}{1998}]{marcon98}Marconi, G., Buonanno, R., Carretta, E. et al. 1998, MNRAS 293, 479 
\bibitem[\protect\astroncite{Montalban et al.}{2000}]{mdm}Montalban, J., D'Antona, F., Mazzitelli, I., 2000, A\&A, 360, 935
\bibitem[\protect\astroncite{Piotto et al.}{1997}]{piotto97}Piotto,
  G., Cool, A. M., King, I. R. 1997, AJ 113, 1345 
\bibitem[\protect\astroncite{Pulone et al.}{2003}]{pulo03}Pulone, L.,
  De Marchi, G., Covino, S., Paresce, F., 2003, A\&A, 399, 121.
\bibitem[\protect\astroncite{Spitzer}{1987}]{spitzer87} Spitzer, L.,
  1987, ``Dynamical Evolution of Globular Clusters'', Princeton
  Univ. Press.
\bibitem[\protect\astroncite{Stetson}{1987}]{stet87}Stetson, P.,  PASP, 99, 191
  1997, 
\bibitem[\protect\astroncite{Reid \& Gizis}{1998}]{reid98}Reid, I., N. $\&$ Gizis, J., E. 1998, AJ 116, 2929 
\bibitem[\protect\astroncite{Renzini et al.}{1996}]{renz96}Renzini,  A., Bragaglia, A. Ferraro, F. et al. 1996,
 ApJ 465L, 23 
\bibitem[\protect\astroncite{Richer et al.}{2000}]{richer00} Richer, H., B., Hansen, B., Limongi, M.,
 et al. ApJ 529, 318 
\end{thebibliography}
\end{document}